\documentclass[pra,aps,superscriptaddress,nofootinbib,twocolumn]{revtex4-1}

\usepackage{amsfonts}
\usepackage{amssymb}
\usepackage{amsmath}
\usepackage{graphicx}
\usepackage[usenames,dvipsnames]{color}
\usepackage[colorlinks=true,citecolor=blue,linkcolor=magenta]{hyperref}
\usepackage{ulem}
\usepackage{lmodern}
\usepackage{amsthm}

\newcommand{\bra}[1]{\left\langle #1 \right|}
\newcommand{\ket}[1]{\left| #1 \right\rangle}
\newcommand{\proj}[1]{\ket{#1}\bra{#1}}
\newcommand{\overlap}[2]{\left\langle #1 | #2 \right\rangle}
\newcommand{\expect}[2]{\left\langle #2 \left| #1 \right| #2 \right\rangle}
\newcommand{\average}[1]{\left\langle #1 \right\rangle}
\newcommand{\abs}[1]{\left| #1 \right|}

\begin{document}

\preprint{APS/123-QED}

\title{Realizing discrete time crystal in an one-dimensional superconducting qubit chain}

\author{Huikai Xu}
\thanks{H. Xu and J. Zhang contributed equally to this work.}
\affiliation{Beijing Academy of Quantum Information Sciences, Beijing 100193, China}

\author{Jingning Zhang}
\thanks{H. Xu and J. Zhang contributed equally to this work.}
\affiliation{Beijing Academy of Quantum Information Sciences, Beijing 100193, China}

\author{Jiaxiu Han}
\affiliation{Beijing Academy of Quantum Information Sciences, Beijing 100193, China}

\author{Zhiyuan Li}
\affiliation{Beijing Academy of Quantum Information Sciences, Beijing 100193, China}

\author{Guangming Xue}
\affiliation{Beijing Academy of Quantum Information Sciences, Beijing 100193, China}

\author{Weiyang Liu}
\email{liuwy@baqis.ac.cn}
\affiliation{Beijing Academy of Quantum Information Sciences, Beijing 100193, China}

\author{Yirong Jin}
\email{jinyr@baqis.ac.cn}
\affiliation{Beijing Academy of Quantum Information Sciences, Beijing 100193, China}

\author{Haifeng Yu}
\email{hfyu@baqis.ac.cn}
\affiliation{Beijing Academy of Quantum Information Sciences, Beijing 100193, China}

\date{\today}

\begin{abstract}
Floquet engineering, i.e. driving the system with periodic Hamiltonians, not only provides great flexibility in analog quantum simulation, but also supports phase structures of great richness. It has been proposed that Floquet systems can support a discrete time-translation symmetry (TTS) broken phase, dubbed the discrete time crystal (DTC). This proposal, as well as the exotic phase, has attracted tremendous interest among the community of quantum simulation. Here we report the observation of the DTC in an one-dimensional superconducting qubit chain. We experimentally realize long-time stroboscopic quantum dynamics of a periodically driven spin system consisting of 8 transmon qubits, and obtain a lifetime of the DTC order limited by the coherence time of the underlying physical platform. We also explore the crossover between the discrete TTS broken and unbroken phases via various physical signatures. Our work extends the usage of superconducting circuit systems in quantum simulation of many-body physics, and provides an experimental tool for investigating non-equilibrium dynamics and phase structures.

\end{abstract}

\maketitle

\textit{Introduction}. Quantum simulation shows great promise for probing the equilibrium and dynamical properties of various quantum phases of many-body systems, which are essential problems in modern condensed matter physics. In addition to static systems, recent theoretical studies~\cite{khemani2016phase, 2016Absolute} show that periodically driven many-body quantum systems, instead of converging to an infinite-temperature thermal state, can support complicated phase structures. Some of the dynamical phases are characterized by the breaking of the discrete  time-translation symmetry (TTS). In other words, these systems will eventually converge to non-equilibrium steady states with periods being integer multiples of that of the driving Hamiltonian. These phases are dubbed discrete time crystal (DTC), following the same way that phases which spontaneously break the spatial translation symmetry are given the name ``crystal''. Ever since the theoretical discovery of the DTC phases, physicists have been considering to realize these exotic non-equilibrium phases with controllable quantum systems.


The DTC phase is an exotic dynamical phase of matter, emerging from the competition of periodic driving, interaction and disorder. The formation of the DTC phase can be understood as follows. First of all, a periodically driven interacting system possesses discrete TTS. It has been shown that the long-time behavior of such systems can be described by an equivalent of the infinite temperature state~\cite{dalessio2014long-time}, with the same TTS as the periodic Hamiltonian. For the emergence of the DTC phase, it is necessary to prevent the ultimate thermalization. A previous theoretical study has shown that there is ergodicity breaking in the Floquet many-body localization (MBL) systems~\cite{ponte2015many, lazarides2015fate}, which are capable of supporting the DTC phase~\cite{else2016floquet}. 
Following this vein, Ref.~\cite{yao2017discrete} proposed to use Floquet engineering in the trapped-ion system to realize the DTC phase. In the same year, the analog simulation of the DTC order has been demonstrated in the trapped-ion system \cite{zhang2017observation} and the nitrogen-vacancy system~\cite{choi2017observation}.

Aside from the analog approach, i.e., directly engineering the system's Hamiltonian as a whole, the superconducting circuit system provides an alternative circuit-based route to the realization of the DTC order, which is made possible by the unprecedented programmability and scalability. Recently, there is an experiment~\cite{frey2021simulating} on the realization of the DTC order using quantum circuits in universal quantum processors consisting of several tens of superconducting transmon qubits. Up to our knowledge, however, the analog approach to the DTC order has not been realized in this powerful physics platform.


In this work, we experimentally demonstrate the DTC phase using an array of 8 capacitively coupled transmon qubits, attached with labels $Q_i$ ($i=1,\ldots,8$), as shown in Fig.~1(a). The major hindrance to the adoption of the theoretical proposal in Ref.~\cite{yao2017discrete} to our system is that the native Hamiltonian of capacitively coupled transmon qubits possesses ${\rm U}(1)$ symmetry, which does not support the DTC phase. We explore the dynamical phase space of the Floquet system by varying the relative strength of the perturbation and the interaction.

\textit{Floquet system}. Floquet systems are physical systems driven by time-periodic Hamiltonians which satisfy $\hat H(t)=\hat H(t+T)$, with $T$ being the period. Floquet's theorem \cite{1965Solution} states that the evolution operator of a Floquet system, i.e. $\hat U\left(t\right)\equiv\hat{\mathcal T}\exp\left(-i\int_0^{t}\hat H\left(t'\right)dt'\right)$ with $\hat{\mathcal T}$ being the time-ordering operator, can be factorized as follows,
\begin{eqnarray}
\hat U\left(t\right)=\hat P\left(t\right)e^{-i\hat H_Ft},
\end{eqnarray}
where $\hat P(t)$ is a unitary operator that satisfies $\hat P\left(t+T\right)=\hat P\left(t\right)$ and $\hat P(0)=\hat{\mathbb I}$. Note that we set $\hbar = 1$ for notation simplicity. Here $\hat H_F$ is the Floquet Hamiltonian that governs the stroboscopic dynamics at times $nT$ ($n=0,1,2,\ldots$), which can be written in the form of spectral decomposition,
\begin{eqnarray}
\hat H_F = \sum_\alpha \epsilon_\alpha \proj{\phi_\alpha},
\end{eqnarray}
with $\epsilon_\alpha\in\left[0,2\pi/T\right)$ and $\ket{\phi_\alpha}$ being the quasi-energies and the Floquet modes, respectively. With this information, the stroboscopic dynamics of an arbitrary initial state $\ket{\Psi(0)}$ can be written as
\begin{eqnarray}
\ket{\Psi(nT)}=\sum_\alpha C_\alpha e^{-in\epsilon_\alpha T}\ket{\phi_\alpha},
\end{eqnarray}
where the expansion coefficients $C_\alpha=\overlap{\phi_\alpha}{\Psi(0)}$ are determined by the initial state $\ket{\Psi(0)}$. 

\textit{DTC phase}. If a spin-$\frac{1}{2}$ system is driven with a sequence of $\pi$-pulses at time points $nT$, where $T$ is much longer than the duration of the $\pi$-pulse, it is clear that the system will return to its initial state in every $2T$ interval. Obviously, the expectation values of observables will oscillate with $2T$ period. However this quantum dynamics is NOT the time crystalline order, since the period-doubled response would be spoiled by even a tiny perturbation in the driving $\pi$-pulses. In other words, to define a dynamical phase, the dynamical response should be stable under the influence of arbitrary weak local $T$-periodic perturbations. As first proposed in equilibrium systems~\cite{1990Ground, 1997Fault, 2005Quasi}, MBL can provide the preceding stability, which inspires the other two ingredients of the DTC phase~\cite{2016Absolute}, i.e. many-body interaction and on-site disorder. The DTC phase can be realized by the Floquet operators, i.e. the propagators for a single period $T =t_1 +t_2$, of the form \cite{else2016floquet} $\hat U_F = \exp\left(-i\hat H_{\rm MBL}t_2\right) \exp\left(-it_1\sum_{i=1}^{N} \hat{\sigma}_i^x\right)$, with $t_1 \simeq \pi/2$ such that the second part approximately forms an Ising parity operator, with $\hat\sigma_i^\alpha$ ($\alpha = x,y,z$) are the Pauli matrices on the $i$-th qubit. One of the spin models that exhibit the MBL is $\hat H_{\rm MBL}=\sum_i\left(J_i\hat\sigma_i^z\hat\sigma_{i+1}^z+h_i^z\hat\sigma_i^z+h_i^x\hat\sigma_i^x\right)$, with $J_i$ being the interaction strength and $h_i^\alpha$ ($\alpha=x,z$) the components of the local random fields. 

{\it Experiment}. The system of transmon qubits can be modeled as coupled Duffing oscillators, whose frequencies $\omega_i$ can be individually tuned by flux. Thus the free evolution of the system is governed by the following Bose-Hubbard model,
\begin{eqnarray}
\hat H&=&\sum_{i=1}^{N}\left(\omega_i\hat a_i^\dag\hat a_i+\frac{\alpha_i}{2}\hat a_i^\dag\hat a_i^\dag\hat a_i\hat a_i\right)\nonumber\\
&&+\sum_{i=1}^{N-1}g_i\left(\hat a_i^\dag+\hat a_i\right)\left(\hat a_{i+1}^\dag+\hat a_{i+1}\right),
\end{eqnarray}
where $\hat a_i$ ($\hat a_i^\dag$) is the annihilation (creation) operator of the $i$-th transmon qubit with the frequency $\omega_i$ and the anharmonicity $\alpha_i$, and $g_i$ quantifies the nearest-neighbor (NN) coupling strength.

\begin{table*}[htbp]
    \centering
    \begin{tabular}{ccccccccc}
        \toprule
        Parameter & Q1 & Q2  & Q3 & Q4 & Q5 & Q6 & Q7 & Q8 \\
        \hline
        $\omega_i/2\pi$ (GHz) & $5.182$  & $4.648$  & $5.146$  & $4.601$  & $5.093$  & $4.560$ & $5.118$  & $4.616$ \\
        $\alpha_i/2\pi$ (GHz) & $-0.240$  & $-0.240$  & $-0.239$  & $-0.240$   & $-0.239$  & $-0.239$  & $-0.239$  & $-0.242$ \\
        $T_{1}$ ($\mu s$) & $15.04$  & $16.79$  & $14.95$  & $16.97$   & $15.86$  & $15.11$  & $12.27$  & $12.54$ \\
        $T_{\phi}$ ($\mu s$) & $10.03$  & $3.86$  & $18.51$  & $3.30$ & $12.99$  & $2.98$  & $23.27$  & $3.51$ \\
        $g_i/2\pi\,(\rm MHz)$ &  \multicolumn{8}{c}{~~~~~~~$19.7$~~~~~~$19.6$~~~~~$19.1$~~~~~$19.1$~~~~~$19.2$~~~~~$19.3$~~~~~$19.6$~~~~~~}\\
        $J_{z,i}/2\pi\,(\rm MHz)$ &  \multicolumn{8}{c}{~~~~~~$1.0$~~~~~~~~$1.2$~~~~~~~$1.1$~~~~~~~$1.3$~~~~~~~$1.1$~~~~~~~$0.9$~~~~~~~$1.2$~~~~~~}\\
        \toprule
    \end{tabular}
    \caption[Device parameters]{Device parameters. $\omega_i$ denotes the qubit frequency, $\alpha_i$ represents the qubit anharmonicity, $T_{1}$ is the qubit relaxation time, $T_{\phi}$ is the qubit dephasing time, $g_i$ is the coupling strength between nearby qubits, and $J_{z,i}$ represents the effective $ZZ$ coupling strength.}\label{tab:Params}
\end{table*}

To engineer the system Hamiltonian into the form of NN Ising $ZZ$-coupling, the frequencies of the qubits at the working points are biased in a zig-zag pattern (see  Table~\ref{tab:Params}). Thus the detunings between neighboring qubits are much larger than the NN coupling strength, i.e. $\abs{\Delta_i}\gg g_i$ with $\Delta_i\equiv\omega_{i+1}-\omega_i$. Under this circumstance, the effective Hamiltonian reads
\begin{eqnarray}
\hat H_{zz}=\sum_{i=1}^{N-1}J_{z,i}\hat\sigma_i^z\hat\sigma_{i+1}^z,
\end{eqnarray}
in the interacting picture. The effect coupling strengths $J_{z,i}$ are calibrated by Ramsey-like experiment \cite{xu2021}.

We realize Floquet drive by time-periodic Hamiltonian, with period $T \equiv t_1+t_2+t_3$
\begin{eqnarray}
\hat H(t) = \left\{\begin{array}{ll} \hat H_x(t) + \hat H_{zz}, & 0\leq t<t_1 \\
\hat H_{zz}, & t_1\leq t < t_1+t_2\\
\hat H_z, & t_1+t_2 \leq t < T \end{array}\right., \label{eq:protocol}
\end{eqnarray}
where
\begin{eqnarray}
\hat H_x(t) &=& g\left(1-\epsilon\right)\sum_{i=1}^Nf_i(t)\hat\sigma_i^x,\label{eq:hamiltonians}\nonumber\\
\hat H_{zz} &=& \sum_{i=1}^{N-1}J_{z,i}\hat\sigma_i^z\hat\sigma_{i+1}^z,\\
\hat H_z &=& \sum_{i=1}^Nh_{z,i}\hat\sigma_i^z\nonumber,
\end{eqnarray}
where $h_{z,i}$ are the strengths of the local longitudinal fields. Within a driving period, the duration of the first interval $t_1$ and the parameter $g$ satisfy the constraint $g t_1=\frac{\pi}{2}$ and $\epsilon$ quantifies the distortion of the $\pi$-pulses. The unitary operator of the third stage $\hat{U}=\exp{\left(-i\sum_i \phi_i\hat\sigma_i^z\right)}$ where $\phi_i \equiv h_{z,i} t_3$ can be realized by virtual Z gates\cite{PhysRevA.96.022330} which actually do not take extra time.
$\hat H_x(t)$ was realized by driving each qubit with resonantly modulated microwave pulse, respectively. The pulse envelope functions $f_i(t)$ can be chosen as arbitrary smooth functions that satisfy the normalization and boundary conditions, i.e. $1/t_1\int_0^{t_1} f_i(t)dt = 1$ and $f_i(0)=f_i(t_1) = 0$. Without loss of generality, we choose Gaussian functions with the full-width-at-half-maximum $w$ in the interval $\left[\frac{t_1}{2}-2w, \frac{t_1}{2}+2w\right]$ as the pulse shaping functions.
The widths of the pulse-envelope functions $w$ are calibrated by  Rabi oscillation experiment.

{\it Results}. Our intention is to investigate Floquet phases with and without discrete TTS breaking. Now we specify the parameters in the driving protocol in Eqs.~(\ref{eq:protocol}) and (\ref{eq:hamiltonians}). In the distortion-free ($\epsilon=0$) and noninteracting ($J_{z,i}=0$) case, the effect of applying $\hat H_x$ in the duration $\left[0, t_1\right)$ on the system is a global $\pi$-rotation upon the $x$-axis in the Bloch sphere. The disorder is introduced by generating random longitudinal fields. Instead of adjusting the Ising interaction strength $J_{z,i}$, which involves tuning the flux bias of each qubit thus increasing the amount of effort in calibration, we fix the interaction strength and vary the duration of the second interval $t_2$, to effectively tune the strength of the interaction and realize different Floquet phases.   

\begin{figure}
  \includegraphics[width=0.45\textwidth]{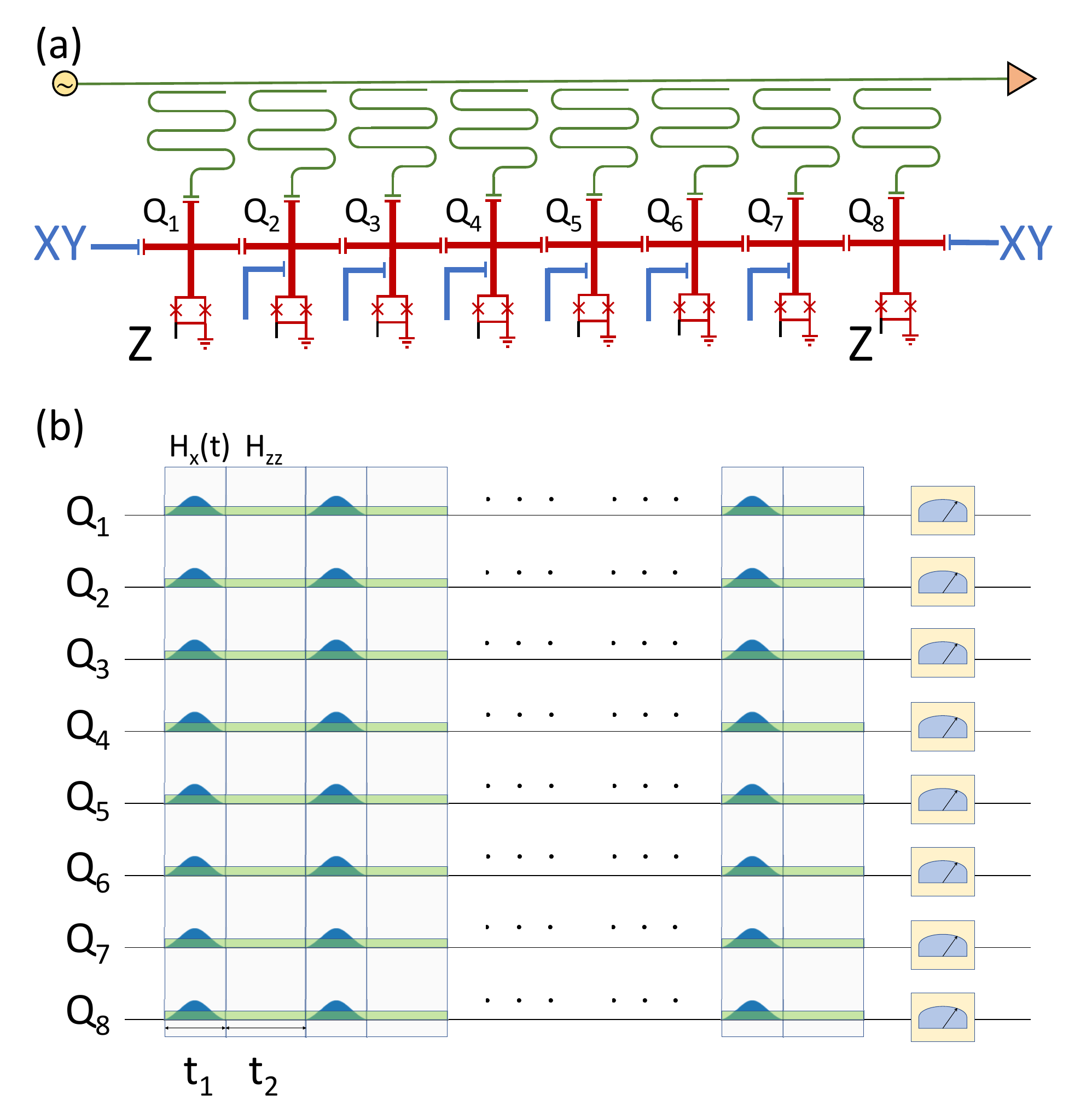}\\
  \caption{Binary driving protocol and experimental setup. (a) Schematic of the 8-transmon device. (b) Schematic of the binary driving protocol in Eq.~(\ref{eq:protocol}). The system is constantly under the influence of NN $ZZ$-coupling, denoted as the green shade. Each period is divided into two intervals with durations $t_1$ and $t_2$. In the first interval, each qubit is driven by a Gaussian-like pulse, shown as the blue wave packets. After driving the system for a number of periods, we measure the qubits to obtain the magnetization.  }\label{fig:Fig1}
\end{figure} 

\begin{figure}
\centering
\includegraphics[width=0.5\textwidth]{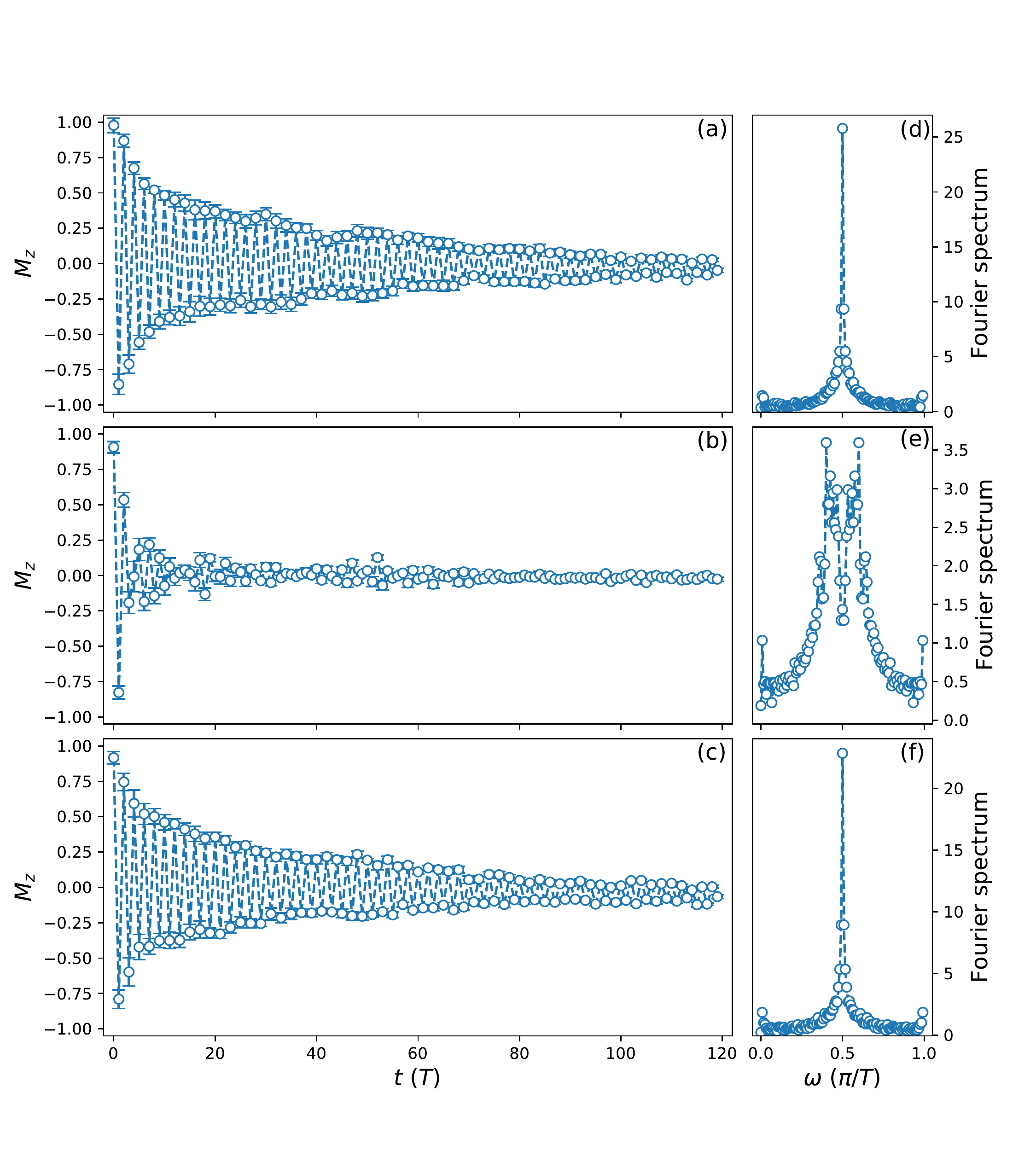}\\
\caption{Representative quantum dynamics in the discrete TTS broken and unbroken phases. (a-c) The stroboscopic dynamics of the magnetization for (a) $\epsilon = 0$, $t_2 = 0$, (b) $\epsilon = 0.18$, $t_2 = 0$, (c) $\epsilon = 0.18$, $t_2 = 50$ ns. Each point in the dynamics is obtained by averaging the expectation values of $\hat\sigma_i^z$, i.e. $\average{\hat\sigma_i^z}$, over 8 qubits, with the error bar showing the standard error of the mean. In our experiment, we repeat each measurement 4096 times to get the expectation values. (d-f) Fourier spectra of the time series in (a-c), respectively. The prominent $\pi/T$-peaks in (d) and (f) clearly show the persistent $2T$-oscillation, thus indicate the breaking of the discrete TTS. }\label{fig:timecrystal}
\end{figure}

To realize different Floquet phases, we fix the duration of the first interval $t_1$ to be 70 ns, and vary the $\pi$-pulse distortion $\epsilon$ and the duration of the second interval $t_2$. The experimental results of representative quantum dynamics in different Floquet phases are summarized in Fig.~2. Firstly, we set the distortion $\epsilon = 0$ and totally suppress the second interval, i.e. $t_2=0$. Due to the existence of strong $ZZ$-coupling, the global $\pi$-pulses cannot be perfect even with vanishing distortion. Fig.~1 (a) shows the $2T$-period oscillation of the magnetization $M_z$, which is defined as
\begin{eqnarray}
M_z\left(t\right)=\frac{1}{N}\expect{\sum_{i=1}^N\hat\sigma_i^z}{\Psi\left(t\right)},
\end{eqnarray}
starting from the fully polarized initial state, i.e. $\ket{\Psi\left(0\right)}=\ket{0\ldots 0}$. Although the oscillation decays under the influence of decoherence, the amplitude is still above the noise level of instrumentation after 120 periods of driving, as shown in Fig.~2 (a).  The Fourier spectrum in Fig.~2 (b) shows a clear peak in $\pi/T$, indicating persistent oscillation with period $2T$. Then we increase the distortion to $\epsilon = 0.18$ and observe that the $2T$-oscillation quickly vanishes and the system converges to a steady state possessing the same symmetry as the driving Hamiltonian, as shown in Fig.~2 (c). The Fourier spectrum (Fig.~2 (d)) is distinct from the previous case by the disappearance of the $\pi/T$-peak. In order to restore the $2T$ oscillation, we increase the duration of the second interval to $t_2=50$ ns. As shown in Fig.~2 (c), the oscillation reappears and persists to at least $120$ periods, even with a longer period $T = 120$ ns. We also find the sharp $\pi/T$-peak reappears in the Fourier spectrum, as shown in Fig.~2 (f).

\begin{figure}[htb]
\centering
\includegraphics[width=0.5\textwidth]{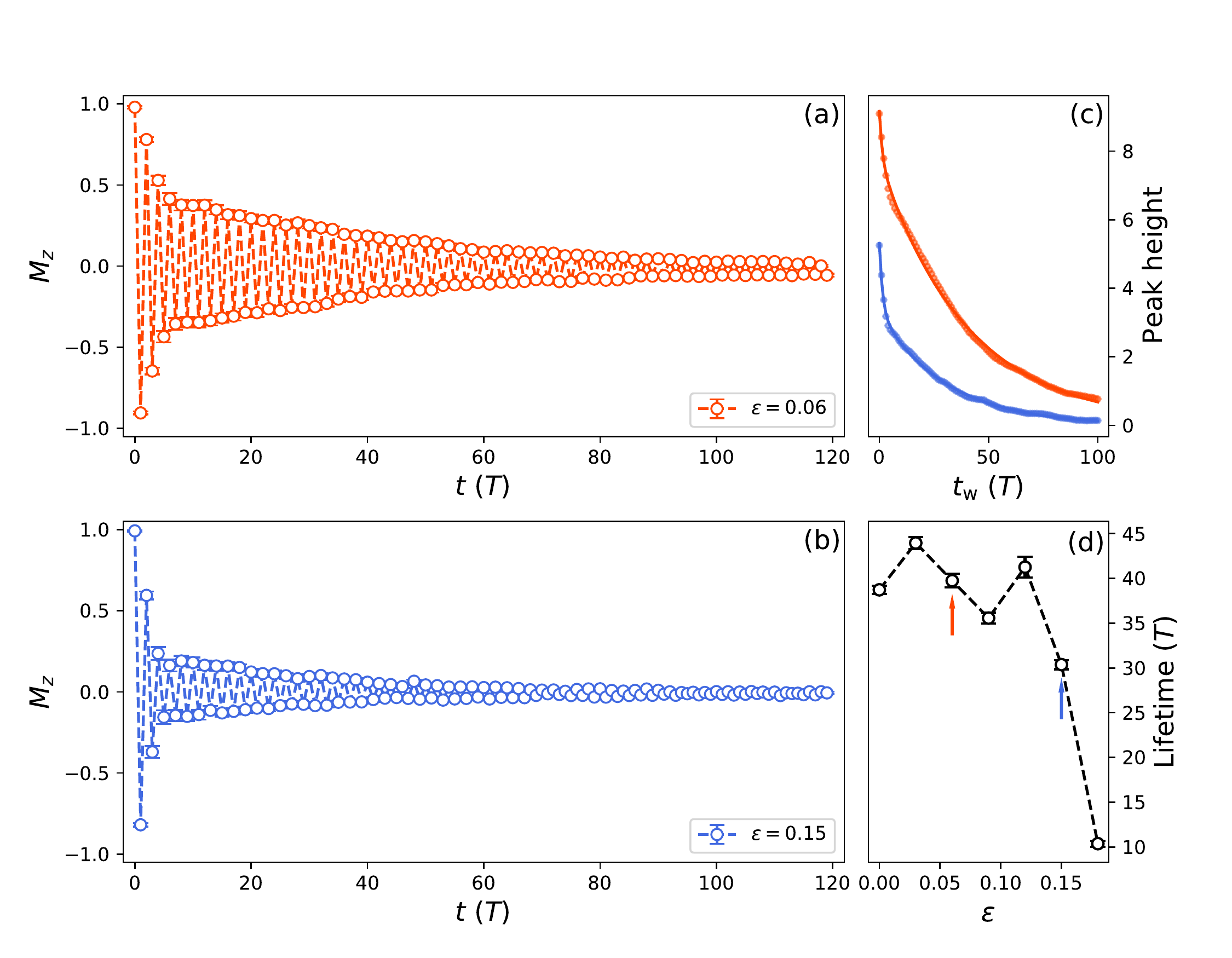}\\
\caption{\text{Lifetime of the DTC order.} The stroboscopic quantum dynamics of the magnetization for (a) $\epsilon = 0.06$, $t_2 = 25$ ns (red) and (b) $\epsilon = 0.15$, $t_2 = 25$ ns (blue). (c) The heights of the $\pi/T$-peaks in the Fourier spectrum of the magnetization in time windows $\left[t_w, t_w+d_w\right]$, with the duration of the window being $d_w = 20 T$, as a function of the position of the starting times $t_w$. The points are from the Fourier transform of the experimental data and the solid lines denotes the bi-exponential fits, with their colors consistent with (a) and (b). (d) Lifetime of the DTC order as a function of the $\pi$-pulse distortion $\epsilon$, with $t_2$ fixed at $25$ ns. The data points corresponding to the experiments in (a) and (b) are marked with arrows. The error bars show the standard error of the mean  over 10 random disorder realizations.}\label{fig:timecrystal}
\end{figure}

\begin{figure}[htb]
  \includegraphics[width=0.5\textwidth]{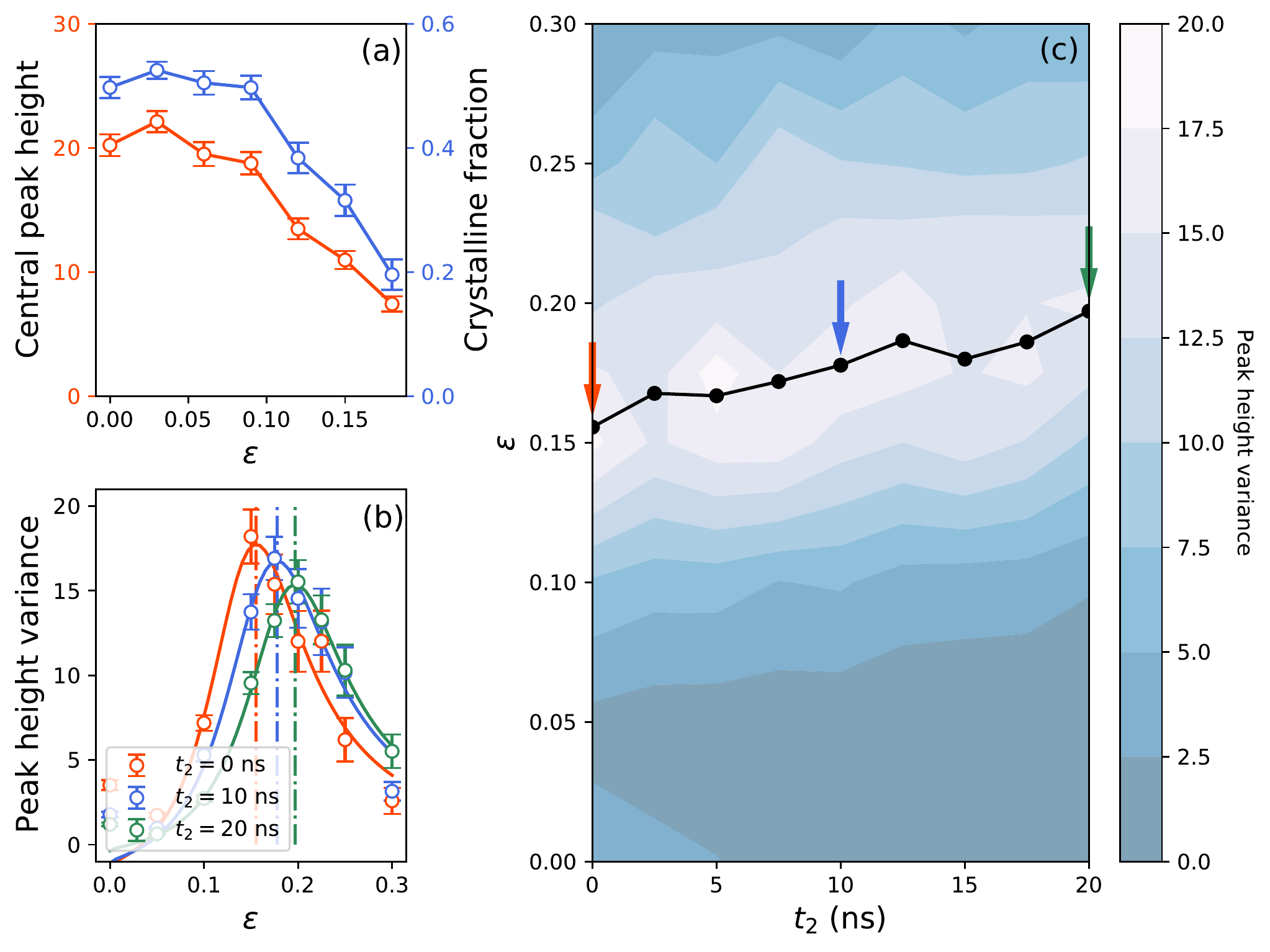}\\
  \caption{Crossover between the discrete TTS broken and unbroken phases. (a) The $\pi/T$-peak heights (red) and the crystalline fraction (blue) as functions of the $\pi$-pulse distortion $\epsilon$. The control parameter $t_2$ is fixed at $25$ ns. The experimental data is accumulated for $100$ disorder realizations and the error bars denote the standard error of the mean obtained by bootstrapping. (b) Variance of the $\pi/T$-peak heights, obtained from exact integration of the Schr$\ddot{\mathrm{o}}$dinger equation, for different values of $t_2$. The peak-height variations are calculated over 100 random disorder realizations, and the error bars are obtained via bootstrapping. The curves are determined by fitting the data points with a Lorenzian-type function $f(\epsilon)=A\left[1+\left(\log\frac{\epsilon}{\epsilon_0}/\gamma\right)^2\right]+B$, with $\epsilon_0$, highlighted by the dashed lines, being the critical points. (c) Phase boundary between the TTS broken and unbroken phases, determined by numerical results of the peak-height variance. The data points corresponding to the curves in (b) are marked with arrows of the same colors.}
\label{fig:crystalinefraction}
\end{figure}
Having confirmed that a discrete TTS breaking phase exists in the bounded regime near $\epsilon = 0$ and finite $t_2$, we further study the robustness of the phase by extracting the lifetime of the persistent oscillation\cite{choi2017observation}. Specifically, we fix $t_2=25$ ns and increase the distortion $\epsilon$ from 0 to 0.18. In Figs.~3 (a) and (b), we show two stroboscopic quantum dynamics of the magnetization with $\epsilon = 0.06$ and $\epsilon = 0.15$. It is clear that larger distortion results in a shorter lifetime of the period-doubled oscillation. To get quantitative information, we set a time window of $20$ periods, move the window along the time axis of the whole dynamics, and calculate the Fourier spectrum in the time window. The heights of the $\pi/T$-peaks for different time windows, with the starting point denoted as $t_w$, are shown in Fig.~3 (c), which take the form of a transient fast decay followed by a much slower one. Inspired by this observation, we fit the $\pi/T$-peak heights with a bi-exponential function $f(t)=\sum_{i=1,2}C_i\exp\left(-t/\tau_i\right)+C_0$, and define the lifetime of the DTC order as ${\rm max}(\tau_i)$. Fig.~3 (d) shows the lifetime of the DTC order as a function of the distortion $\epsilon$. The lifetime keeps almost constant with $\epsilon\leq 0.12$ for $t_2 = 25$ ns, and then it starts to decrease. In the DTC phase, the lifetime is presumably limited by the finite coherence times of the transmon qubits.  

Finally, we confirm the crossover of the discrete TTS broken and unbroken phases with more physical signatures. Considering the Fourier spectrum of the whole stroboscopic dynamics of the magnetization, we show the heights of the $\pi/T$-peak and the crystalline fraction, defined as the ratio of the $\pi/T$-peak height to the sum of the whole spectrum, as functions of the distortion $\epsilon$ with $t_2 = 25$ ns in Fig.~4 (a), respectively. They both show similar behavior as the lifetime of the DTC order. Note that the phase boundary between these phases are blurry due to the finite size of our experimental system. We've tried the proposal in Ref.~\cite{yao2017discrete}, which determines the phase boundary by finding the maximum of the peak-height variance among different disorder realizations. The reason that we didn't find a clear phase boundary may be due to the instability of the experimental device during the course of data accumulation, which takes several days for the case of 100 disorder realizations for each set of control parameters. 

\textit{Conclusion}. We experimentally realize the discrete TTS broken and unbroken phases in a periodically driven superconducting circuit system, and explore the crossover between these phases. Our experiment shows the feasibility of using Floquet engineering and superconducting circuit systems to investigate non-equilibrium dynamics and dynamical phase structure, which is of essential importance in the frontier of condensed matter physics. Thanks to the controllability and scalability of the superconducting quantum circuits, it is feasible to extend our experiment to larger systems, even beyond the limit that classical computers can efficiently handle. Such experimental protocols also pose challenges on practical techniques, calling for longer coherence times and better stability, due to the fact that a single experiment, as well as the data accumulation phase, lasts for a relatively long time.

\textit{Note}. During the preparation of this paper, we found a similar work by the Google Quantum AI team. They used digital simulation to achieve DTC in superconducting quantum circuits\cite{google2021}.

\textit{Acknowledgments}. We appreciate the helpful discussion with Yingshan Zhang and Ruixia Wang. This work was supported by the NSF of Beijing (Grant No.Z190012), the NSFC of China (Grants No. 11890704, No.12004042,No. 11905100), National Key Research and Development Program of China (Grant No.2016YFA0301800),and the Key-Area Research and Development Program of
Guang Dong Province (Grant No. 2018B030326001).


\begin{thebibliography}{10}

\bibitem{khemani2016phase}
Vedika Khemani, Achilleas Lazarides, Roderich Moessner, and S.~L. Sondhi.
\newblock Phase structure of driven quantum systems.
\newblock {\em Phys. Rev. Lett.}, 116:250401, Jun 2016.

\bibitem{2016Absolute}
C.~W. von Keyserlingk, Vedika Khemani, and S.~L. Sondhi.
\newblock Absolute stability and spatiotemporal long-range order in floquet
  systems.
\newblock {\em Phys. Rev. B}, 94:085112, Aug 2016.

\bibitem{dalessio2014long-time}
Luca D'Alessio and Marcos Rigol.
\newblock Long-time behavior of isolated periodically driven interacting
  lattice systems.
\newblock {\em Phys. Rev. X}, 4:041048, Dec 2014.

\bibitem{ponte2015many}
Pedro Ponte, Z.~Papi\ifmmode~\acute{c}\else \'{c}\fi{}, Fran\ifmmode
  \mbox{\c{c}}\else~\c{c}\fi{}ois Huveneers, and Dmitry~A. Abanin.
\newblock Many-body localization in periodically driven systems.
\newblock {\em Phys. Rev. Lett.}, 114:140401, Apr 2015.

\bibitem{lazarides2015fate}
Achilleas Lazarides, Arnab Das, and Roderich Moessner.
\newblock Fate of many-body localization under periodic driving.
\newblock {\em Phys. Rev. Lett.}, 115:030402, Jul 2015.

\bibitem{else2016floquet}
Dominic~V. Else, Bela Bauer, and Chetan Nayak.
\newblock Floquet time crystals.
\newblock {\em Phys. Rev. Lett.}, 117:090402, Aug 2016.

\bibitem{yao2017discrete}
N.~Y. Yao, A.~C. Potter, I.-D. Potirniche, and A.~Vishwanath.
\newblock Discrete time crystals: Rigidity, criticality, and realizations.
\newblock {\em Phys. Rev. Lett.}, 118:030401, Jan 2017.

\bibitem{zhang2017observation}
J.~Zhang, P.~W. Hess, A.~Kyprianidis, P.~Becker, A.~Lee, J.~Smith, G.~Pagano,
  I.-D. Potirniche, A.~C. Potter, A.~Vishwanath, N.~Y. Yao, and C.~Monroe.
\newblock Observation of a discrete time crystal.
\newblock {\em Nature}, 543(Mar.9 TN.7644):217--220, 2017.

\bibitem{choi2017observation}
S.~Choi, J.~Choi, R.~Landig, G.~Kucsko, H.~Zhou, J.~Isoya, Jelezko F.,
  S.~Onoda, H.~Sumiya, V.~Khemani, Keyserlingk C., Norman Y., Demler E., and
  Lukin M.
\newblock Observation of discrete time-crystalline order in a disordered
  dipolar many-body system.
\newblock {\em Nature}, 543:221, 2017.

\bibitem{frey2021simulating}
P.~Frey and S.~Rachel.
\newblock Simulating a discrete time crystal over 57 qubits on a quantum
  computer.
\newblock {\em arXiv:2105.06632}.

\bibitem{1965Solution}
J.~H. Shirley.
\newblock Solution of the schrdinger equation with a hamiltonian periodic in
  time.
\newblock {\em Physical Review}, 138(4B):979--987, 1965.

\bibitem{1990Ground}
X.~G. Wen and Q.~Niu.
\newblock Ground-state degeneracy of the fractional quantum hall states in the
  presence of a random potential and on high-genus riemann surfaces.
\newblock {\em Phys. Rev. B}, 41:9377--9396, May 1990.

\bibitem{1997Fault}
A.Yu. Kitaev.
\newblock Fault-tolerant quantum computation by anyons.
\newblock {\em Annals of Physics}, 303(1):2--30, 2003.

\bibitem{2005Quasi}
M.~B. Hastings and Xiao-Gang Wen.
\newblock Quasiadiabatic continuation of quantum states: The stability of
  topological ground-state degeneracy and emergent gauge invariance.
\newblock {\em Phys. Rev. B}, 72:045141, Jul 2005.

\bibitem{xu2021}
Zhiyuan Li Jiaxiu Han Jingning Zhang Kehuan Linghu Yongchao Li Mo Chen Zhen
  Yang Junhua Wang Teng Ma Guangming Xue Yirong~Jin Huikai~Xu, Weiyang~Liu and
  Haifeng Yu.
\newblock Realization of adiabatic and diabatic cz gates in superconducting
  qubits coupled with a tunable coupler.
\newblock {\em Chinese Physics B}, 30(4):044212, 2021.

\bibitem{PhysRevA.96.022330}
David~C. McKay, Christopher~J. Wood, Sarah Sheldon, Jerry~M. Chow, and Jay~M.
  Gambetta.
\newblock Efficient $z$ gates for quantum computing.
\newblock {\em Phys. Rev. A}, 96:022330, Aug 2017.

\bibitem{google2021}
Google~Quantum AI and collaborators.
\newblock Observation of time-crystalline eigenstate order on a quantum
  processor.
\newblock {\em arXiv:2107.13571}, 2021.

\end{thebibliography}
\end{document}